\documentclass[pdflatex,sn-mathphys-num]{sn-jnl}

\usepackage{graphicx}%
\usepackage{multirow}%
\usepackage{amsmath,amssymb,amsfonts}%
\usepackage{amsthm}%
\usepackage{mathrsfs}%
\usepackage[title]{appendix}%
\usepackage{xcolor}%
\usepackage{textcomp}%
\usepackage{manyfoot}%
\usepackage{booktabs}%
\usepackage{algorithm}%
\usepackage{algorithmicx}%
\usepackage{algpseudocode}%
\usepackage{listings}%
\usepackage{hyperref}%

\theoremstyle{thmstyleone}%
\theoremstyle{thmstyletwo}%

\theoremstyle{thmstylethree}%

\begin{document}

\title[Geometric structure of two-neutron halo nuclei from Efimov
physics at the unitary limit]{Geometric structure of two-neutron halo nuclei from Efimov
physics at the unitary limit}


\author*[1,2]{\fnm{R.} \sur{M. Francisco}}\email{rafaelrmf@ita.br}\equalcont{These authors contributed equally to this work.}

\author[1]{\fnm{D.} \sur{S. Rosa}}\email{derickdsr@ita.br
}
\equalcont{These authors contributed equally to this work.}

\author*[1]{\fnm{T.} \sur{Frederico}}\email{tobias@ita.br}
\equalcont{These authors contributed equally to this work.}

\author[3]{\fnm{M.} \sur{T. Yamashita}}\email{marcelo.yamashita@unesp.br}\equalcont{These authors contributed equally to this work.}

\affil*[1]{\orgname{Instituto Tecnol\'{o}gico de Aeron\'{a}utica}, \orgdiv{DCTA},\orgaddress{  \postcode{12228-900}, \city{S\~{a}o Jos\'{e} dos Campos}, \country{Brazil}}}

\affil*[2]{\orgname{Universit\'e Paris-Saclay}, \orgdiv{CNRS/IN2P3, IJCLab}, \orgaddress{\postcode{91405}, \city{Orsay}, \country{France}}}

\affil[3]{\orgdiv{Instituto de F\'isica Te\'orica}, \orgname{Universidade Estadual Paulista}, \orgaddress{\street{Rua Dr. Bento Teobaldo Ferraz, 271-Bloco II}, \city{S\~ao Paulo}, \postcode{01140-070}, \state{SP}, \country{Brazil}}}


\abstract{We investigate the geometric structure of two-neutron halo nuclei from the perspective of Efimov physics. Using the analytic three-body wave function obtained from the Faddeev equations in the unitary limit, we explore the connection between Efimov universality and the spatial configuration of these weakly bound systems. The internal geometry is quantified through probability densities, root-mean-square interparticle distances and characteristic opening angles, evaluated for different neutron–core mass ratios. Our results reveal a universal trend in the geometry of $s$-wave dominated halo nuclei, reflecting the universal correlations characteristic of the Efimov-like regime.}




\maketitle

\section{Introduction}\label{sec1}

Halo nuclei represent one of the most fascinating manifestations of quantum few-body dynamics. In light exotic nuclei such as $^{11}$Li, $^{14}$Be, $^{19}$B, and $^{22}$C, two weakly bound neutrons form a diffuse cloud surrounding a compact core~\cite{Hansen1987,AlKhalili1996}. These systems exhibit extended spatial distributions and extremely low binding energies, placing them near the breakup threshold. Under these conditions, the neutron–neutron and neutron–core subsystems often correspond to virtual or weakly bound states close to zero energy~\cite{Fedorov1994,Amorim1997,Hammer2022,Naidon2023}. When this occurs, the halo structure can display universal properties largely independent of the details of the nuclear interactions, making these nuclei a valuable testing ground for few-body universality in the nuclear domain~\cite{Frederico2012,Bertulani2002,Canham2008,Canham2010}.

The theoretical foundation underlying this universality traces back to Efimov’s seminal prediction in 1970~\cite{Efimov1970}. While studying the quantum three-body problem with short-range interactions, Efimov discovered a remarkable scaling law for systems where the two-body subsystems are at or near the unitary limit. He showed that, for three identical bosons interacting in the $s$-wave channel, the three-body spectrum exhibits a geometric sequence of bound states whose energies satisfy a constant ratio between successive levels. As the two-body binding energy approaches zero, these Efimov states accumulate near the three-body threshold. For finite two-body binding energies, however, only a finite number of Efimov states survive~\cite{Efimov1991,Rosa2018}. These features: discrete scale invariance, geometric scaling and universality, constitute what is now known as Efimov physics (for general results on Efimov physics see Refs.~\cite{Braaten2006,Francisco2022,Francisco2025,Naidon2017}).

The connection between Efimov physics and halo nuclei arises naturally from their underlying few-body character and from the dominance of $s$-wave interactions in their subsystems. In the limit of weak binding and short-range forces, the structural and dynamical properties of two-neutron halo nuclei, where both the neutron–neutron and neutron–core interactions occur predominantly in $s$-wave channels, may reflect the same universal correlations predicted by Efimov. In realistic halo nuclei, however, the finite interaction ranges, finite scattering lengths, and limited separation of scales usually prevent the appearance of a full Efimov tower, so that at most the ground-state Efimov-like configuration is expected to survive. Although the strict Efimov scaling hierarchy is not expected to manifest fully in these nuclear systems, the spatial configurations of these halos can still encode the geometric signatures characteristic of Efimov-like states.

In this contribution, we explore these aspects by analyzing how Efimov-like structures emerge in mass-imbalanced three-body systems relevant to two-neutron halo nuclei. Our approach relies on the $D = 3$ form of the three-body $D$-dimensional wave function at unitarity, derived through the Bethe–Peierls boundary condition~\cite{Rosa2022}. For three identical bosons, this formulation reproduces the well-known Efimov wave function presented in Refs.~\cite{Efimov1971,Castin2011}. In the case of halo nuclei, we consider the s-wave singlet neutron-neutron channel where the spatial wave function is symmetric and include the appropriate mass-dependent weights in the Faddeev components to account for the neutron–core mass imbalance. From this framework, we use analytical expressions for key spatial observables—such as the mean separation distances between the constituents, opening angles, and density distributions. By examining how these quantities depend on the mass ratio, our goal is to explore the spatial structure of two-neutron halo systems in the unitary limit and exemplify the extent to which Efimov-type universality governs their geometry.

Closely related is the recent work of Naidon~\cite{Naidon2023}, who studied
the universal geometry of two-neutron Borromean halos within a Faddeev-based
framework, including some of the same halo nuclei considered in the present
work. In contrast to that study, which
retains finite two-body input and treats configurations close to dissociation,
we work at the strict unitary limit, where the three-body wave function is
fully analytic and the two-neutron separation energy is the only scale; this
yields closed-form expressions for all geometric observables---distances,
opening angles, and densities---as functions of the neutron--core mass ratio.
Our unitary results thus provide a parameter-free baseline against which
finite-range corrections can be measured, as illustrated by the direct
$^{11}$Li comparison in Section~\ref{subsec3.2}.

\section{Brief Review of the Formalism}\label{sec2}

In this section, we review the formalism we use to study the two-neutron halo nuclei at unitarity. For a complete derivation of the eigenstates of the Hamiltonian for contact interactions, see Ref.~\cite{Naidon2017}.
In what follows, we present the derivation of the eigenstates of the contact interaction Hamiltonian in $D-$dimensions~\cite{Rosa2022}, which has application in squeezed trapped cold atoms close to a Feshbach resonance to analyze theoretically the transition from the discrete scaling  symmetry to the continuous one or unatomic regime~\cite{Francisco2025}. 

In the derivation of the Hamiltonian eigenstates it is used the Faddeev decomposition of the wave function alongside with the Jacobi coordinates 
\begin{align}
    \vec{r}_i = \vec{x}_j - \vec{x}_k\quad 
    \text{and}\quad\vec{\rho}_i = \vec{x}_i - \frac{m_j \vec{x}_j + m_k \vec{x}_k}{m_j + m_k},
    \label{Jacobic}
\end{align}
where $\vec{x}_i$, $\vec{x}_j$ and $\vec{x}_k$ are the vector positions of each particle in the three-body system, while $m_j$ and $m_k$ are the masses of particles $j$ and $k$. $\vec{r}_i$ describes the relative position between two particles (e.g., a neutron--neutron pair) and $\vec{\rho}_i$ describes the relative position between the third particle and the center of mass of the other two.

To symmetrize the kinetic-energy operator in the Hamiltonian, we scale the Jacobi coordinates as $\vec{r}'_{i} =  \sqrt{\mu_{jk}} \vec{r}_{i}$ and $\vec{\rho}'_{i} = \sqrt{\mu_{jk,i}} \vec{\rho}_{i}$, where $\mu_{jk} = \frac{m_j m_k}{m_j + m_k}$ and 
$\mu_{jk,i} = \frac{m_i (m_j + m_k)}{m_i + m_j + m_k}$. The transformations between different pairs of scaled Jacobi coordinates read
\begin{align}
 \vec{r}\,'_{i}& = - \vec{r}\,'_{j}\cos\phi_k + \vec{\rho}\,'_{j}\sin\phi_k \nonumber \\
  \vec{\rho}\,'_{i}& = - \vec{r}\,'_{j}\sin\phi_k - \vec{\rho}\,'_{j}\cos\phi_k\,,
  \label{transf}
\end{align}
where the angles that describe the transformation are given by
\begin{align}
 \phi_k = \arctan \sqrt{\frac{m_{k} (m_i + m_j + m_k)}{m_i m_j}}\,.
 \label{transformationang}
\end{align}
 
Introducing the hyperspherical coordinates, $R^2 = |\vec{r}_i\,'|^2 + |\vec{\rho}_i\,'|^2$ and $\alpha_i = \arctan\left(\frac{|\vec{r}_i\,'|}{|\vec{\rho}_i\,'|}\right)$, the D-dimensional wave function is expressed as a hyperradial factor multiplying the sum of the three hyperangular Faddeev components:
\begin{align}
   \Psi(R,\alpha_i) = F_{\mathrm{i}s_0}(R)\sum_{i = 1}^{3}G(\alpha_i)\quad \text{with}\quad F_{\mathrm{i} s_0}(R) = \frac{K_{ \mathrm{i}s_0}\left(\kappa_0 \sqrt{2} R \right) }{R^{D-1}}\,,
\end{align}
and
\begin{multline}
  G(\alpha_i) =  C^{(i)}\frac{(\sin2 \alpha_i)^{1/2}}{(\sin\alpha_i\cos\alpha_i)^{(D-1)/2}}  \left[P_{\mathrm{i} s_0/2-1/2}^{D/2-1}(\cos2 \alpha_i)\right.  \\
 -
 \left.\frac{2}{\pi}\tan\left[\pi\left(\frac{\mathrm{i}s_0-1}{2} \right)\right] Q_{\mathrm{i}s_0/2-1/2}^{D/2-1}(\cos 2 \alpha_i)\right],
 \label{angularpart}
\end{multline}
 where $K_{\mathrm{i} s_0}$ is the modified Bessel function of the third kind of pure imaginary order, $\mathrm{i}s_0$, $P_{\mathrm{i} s_0/2-1/2}^{D/2-1}(\cos2 \alpha_i)$ and $Q_{\mathrm{i}s_0/2-1/2}^{D/2-1}(\cos 2 \alpha_i)$ are the associated Legendre functions and $-\kappa_0^2 = E_3$ in units where $\hbar = m_n = 1$. The $D$ dependence enters both the hyperangular function and the power-law prefactor of the hyperradial part.
For $D=3$, one has $\mu = D/2 - 1 = 1/2$, so that the standard identities~\cite{Abramowitz1964}
\begin{align}
   P_{\mathrm{i} s_0/2 - 1/2}^{1/2}(\cos 2\alpha)
   =& \sqrt{\frac{2}{\pi \sin(2\alpha)}}\, \cos(\mathrm{i} s_0 \alpha) \,,
   \end{align}
   and
   \begin{align}
   Q_{\mathrm{i} s_0/2 - 1/2}^{1/2}(\cos 2\alpha)
   =& -\sqrt{\frac{\pi}{2 \sin(2\alpha)}}\, \sin(\mathrm{i} s_0 \alpha)\,,
\end{align}
can be used to write Eq.~\eqref{angularpart} in the form
\begin{equation}
   G(\alpha) = 
   C^{(i)}\,\frac{2}{\sqrt{\sin(2\alpha)}}\left(\frac{2}{\pi \sin(2\alpha)}\right)^{1/2}
   \left[
      \cos(\mathrm{i} s_0 \alpha) + \tan\!\Big(\frac{\pi(\mathrm{i} s_0 - 1)}{2}\Big)\, \sin(\mathrm{i} s_0 \alpha)
   \right]\,.
   \label{eq:G_cos_sin}
\end{equation}
 Next, one can use the trigonometric identity, $\tan(u - \tfrac{\pi}{2}) = -\cot u$, to rewrite
\begin{align}
   \tan\!\Big(\frac{\pi \mathrm{i} s_0}{2} - \frac{\pi}{2}\Big)
   = -\cot\!\Big(\frac{\pi \mathrm{i} s_0}{2}\Big).
\end{align}
Then, the expression inside the brackets in Eq.~\eqref{eq:G_cos_sin} becomes
\begin{align}
   \cos(\mathrm{i} s_0 \alpha) - \cot\!\Big(\frac{\pi \mathrm{i} s_0}{2}\Big)\sin(\mathrm{i} s_0 \alpha)\,.
\end{align}
 Using the sine subtraction formula
\begin{align}
   \sin\!\big[\mathrm{i} s_0(\tfrac{\pi}{2}-\alpha)\big]
   = \sin\!\Big(\frac{\pi \mathrm{i} s_0}{2}\Big)\cos(\mathrm{i} s_0 \alpha)
     - \cos\!\Big(\frac{\pi \mathrm{i} s_0}{2}\Big)\sin(\mathrm{i} s_0 \alpha)
\end{align}
and dividing it by $\sin(\tfrac{\pi \mathrm{i} s_0}{2})$, one gets
\begin{align}
\frac{\sin\!\big[\mathrm{i} s_0(\tfrac{\pi}{2}-\alpha)\big]}      {\sin\!\big(\tfrac{\pi \mathrm{i} s_0}{2}\big)}
   = \cos(\mathrm{i} s_0 \alpha)
     - \cot\!\Big(\frac{\pi \mathrm{i} s_0}{2}\Big)\sin(\mathrm{i} s_0 \alpha)\,.
\end{align}
Using this result in Eq.~\eqref{eq:G_cos_sin} gives
\begin{equation}
   G(\alpha) =
   C^{(i)}\,\frac{2}{\sqrt{\sin(2\alpha)}}\left(\frac{2}{\pi \sin(2\alpha)}\right)^{1/2}
  \frac{\sin\!\big[\mathrm{i} s_0(\tfrac{\pi}{2}-\alpha)\big]}{\sin\!\big(\tfrac{\pi \mathrm{i} s_0}{2}\big)}\,
   .
\end{equation}
\noindent Absorbing the constant prefactors into the normalization constant and restoring the hyperradial function, one obtains the final
three-body wave function at unitarity in three dimensions~\cite{Bulgac1976,Efimov1971,Castin2011}
\begin{equation}
   \Psi(R,\alpha_i)
   = N_{\Psi}\sum_{i=1}^{3}C^{(i)}
   \frac{K_{ \mathrm{i} s_0}\!\big(\kappa_0 \sqrt{2} R\big)}{R^{2}}
   \frac{
      \sin\!\big[\mathrm{i} s_0\big(\tfrac{\pi}{2}-\alpha_i\big)\big]
   }{
      \sin2\alpha_i}
   ,
   \label{eq:totalwavefunction}
\end{equation}
\noindent where $N_{\Psi}$ is the normalization constant.

\begin{figure}[ht!]
\centering
\includegraphics[width=6.47cm]{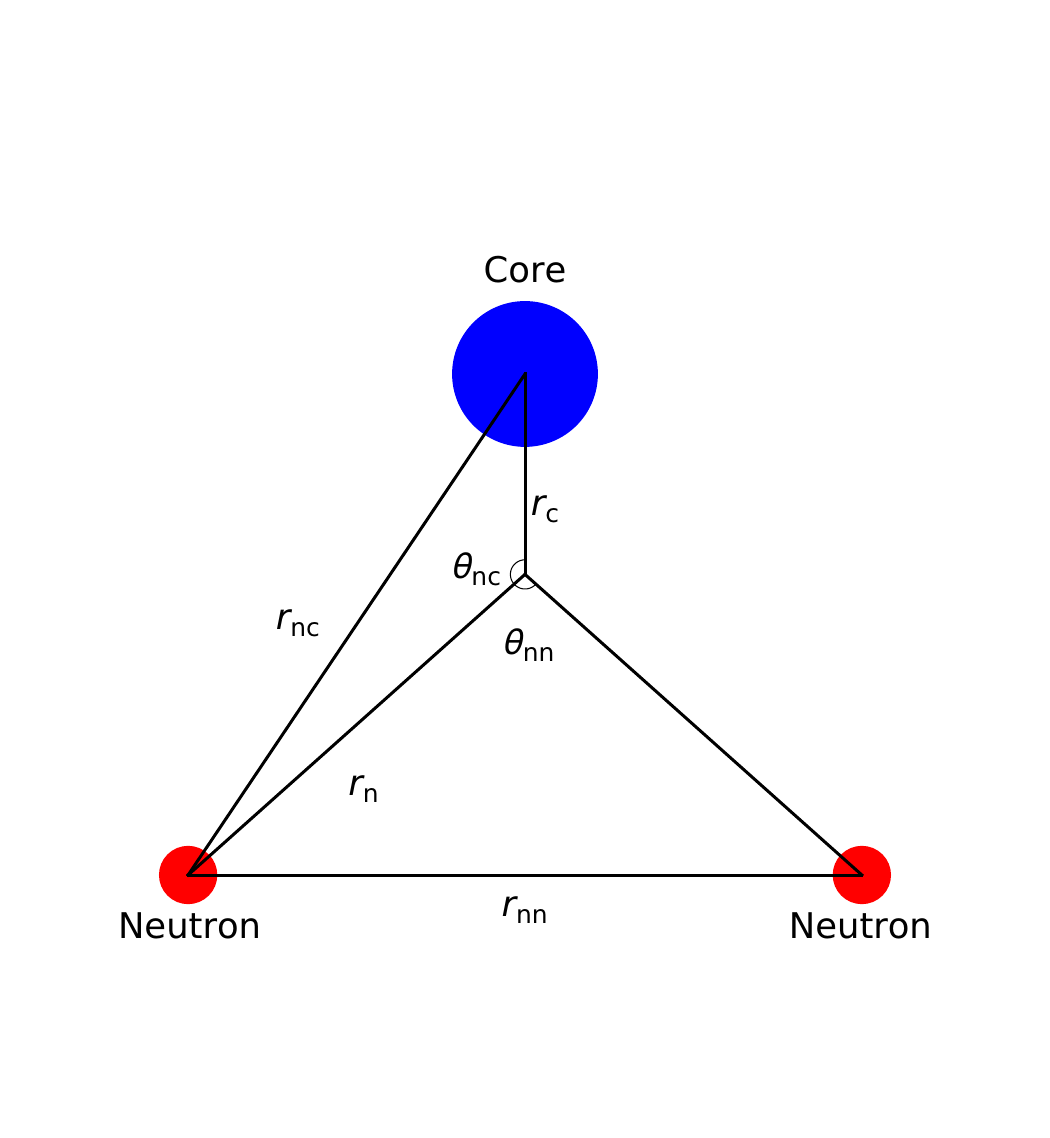}
\centering
\includegraphics[width=6.47cm]{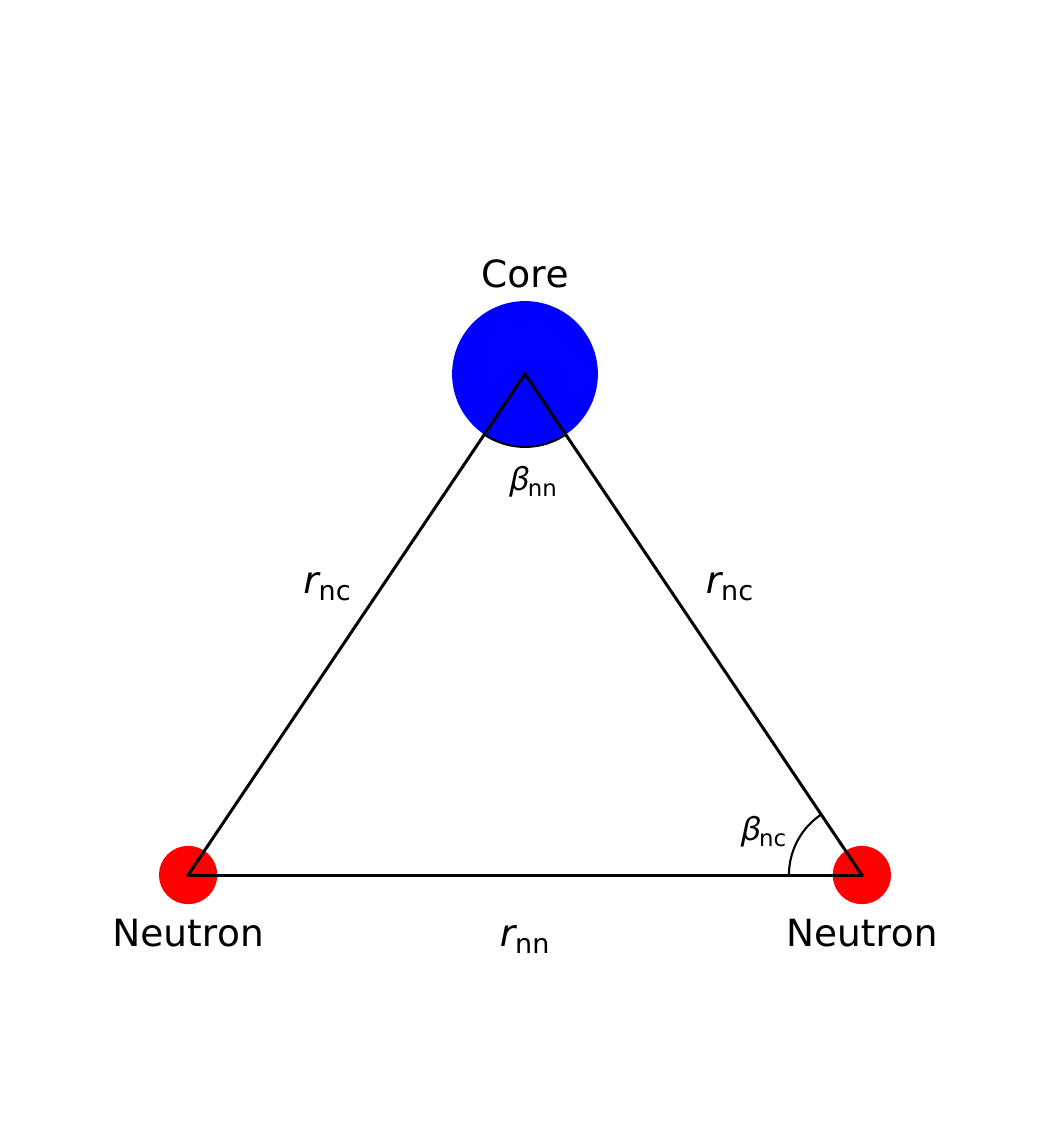}
\vspace{-1cm}
\caption{Illustration of the relative distances and opening angles of the three-body system. The left panel defines opening angles at the
three-body center of mass while the right panel defines them at the individual particle centers.} \label{figangles} 
\end{figure}

The three-body wave function, Eq.~\eqref{eq:totalwavefunction}, can be used to calculate the dimensionless root mean square distances
\begin{align}
\sqrt{\langle O^{2}\rangle\,|E_3|},
\label{dimensioless}
\end{align}
 where $O$ represents any relative distance depicted in Fig.~\ref{figangles}. It is well known that the quantity in Eq.~\eqref{dimensioless} is universal. This can be seen by remembering that the Efimov effect is characterized by a discrete scale invariance: the system exhibits identical physical behavior under the transformations
\begin{align}
    \langle O^2 \rangle \to \lambda^2 \langle O^2 \rangle, \quad E_3 \to \frac{E_3}{\lambda^2}\,,
    \label{observables}
\end{align}
where the scaling factor is given by $\lambda = e^{\pi/s_0}$. 

Using the results obtained with Eq.~\eqref{observables}, the relative opening angles, exhibited in Fig.~\ref{figangles}, can be written as 
\begin{align}
    &\theta_{nc} = \cos^{-1} \frac{\langle \vec{r}_c \cdot \vec{r}_n \rangle}{\sqrt{\langle r_c^2 \rangle \langle r_n^2 \rangle}} = \cos^{-1} \left(\frac{\langle r_c^2 \rangle + \langle r_n^2 \rangle - \langle r_{nc}^2 \rangle}{2 \sqrt{\langle r_c^2 \rangle \langle r_n^2 \rangle}}\right)\,, \nonumber \\
&\theta_{nn} = \cos^{-1} \frac{\langle \vec{r}_n \cdot \vec{r}_{n'} \rangle}{\langle r_n^2 \rangle} = \cos^{-1} \left( 1 - \frac{1}{2} \frac{\langle r_{nn}^2 \rangle}{\langle r_n^2 \rangle} \right)\,,
\end{align}
and 
\begin{align}
&\beta_{nc} = \cos^{-1} \frac{ \langle \vec{r}_{nc} \cdot \vec{r}_{nn'} \rangle }{ \sqrt{ \langle r_{nc}^2 \rangle \langle r_{nn}^2 \rangle } } = \cos^{-1} \left( \frac{\sqrt{\langle r^{2}_{nn} \rangle}}{2 \sqrt{\langle r^2_{nc} \rangle}} \right)   \nonumber \\
&\beta_{nn} = \cos^{-1} \frac{ \langle \vec{r}_{nc} \cdot \vec{r}_{n'c} \rangle }{  \langle r_{nc}^2 \rangle  }  = \cos^{-1}\left(1 - \frac{\langle r_{nn}^2 \rangle}{2 \langle r_{nc}^2 \rangle}\right)\,.
\end{align}
With the above equations, we can study the behavior of two-neutron halo nuclei in the approximation where the two-body interactions are taken to be at the unitary limit. In the next section, we will explore the geometry of these systems by calculating some distributions, besides root mean square values for relative distances and opening angles.

\section{Results and discussion}\label{sec3}

In this section, we investigate the geometry of halo nuclei systems at the unitary limit. We use the analytical three-body wave function at the unitary limit to obtain results for probability densities, average interparticle distances and opening angles. For completeness, we show results for both $A \le 1$ and $A > 1$, where $A = \frac{m_c}{m_n}$, being $m_c$ the mass of the core and $m_n = 1$ the mass of the neutrons. 

\subsection{Efimov scale parameter and probability densities}\label{subsec3.1}

We begin by presenting the Efimov scaling parameter $s_0$ and the ratio between the Faddeev-component weights as functions of the mass ratio $A$ for a YYZ system. In this configuration, two particles (Y) have equal masses while the third particle (Z) has a different mass, which corresponds to the structure of two-neutron halo nuclei.

To describe this system we use the three-body wave function defined in Eq.~\eqref{totalwav1}, expressed in T-type Jacobi coordinates. In this coordinate set, the variable $|\vec{r}_z'|\equiv r$ represents the distance between the two identical particles (YY), while $|\vec{\rho}_z'|$ measures the distance between the third particle (Z) and the center of mass of the YY subsystem. With this choice of coordinates, the coefficient $C^{(z)}$ gives the weight of the Faddeev component written in the Z (or T) Jacobi set, whereas $C^{(y)}$ corresponds to the components associated with the Y Jacobi sets. The latter are obtained by expressing the Y Jacobi coordinates in terms of the Z coordinate set.

\begin{figure}[ht!]
\centering
\includegraphics[width=6.4cm]{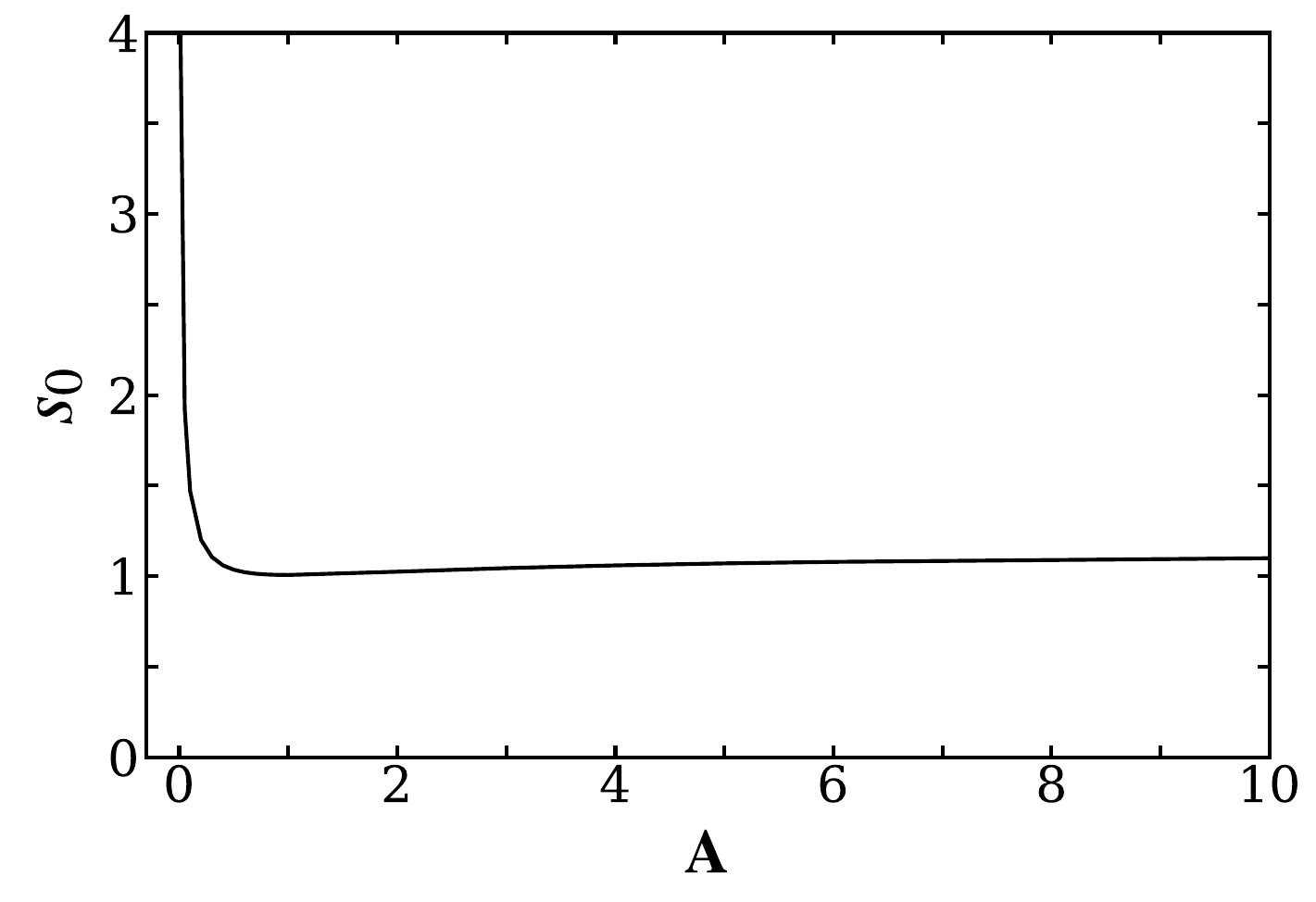} 
\centering
\includegraphics[width=6.5cm]{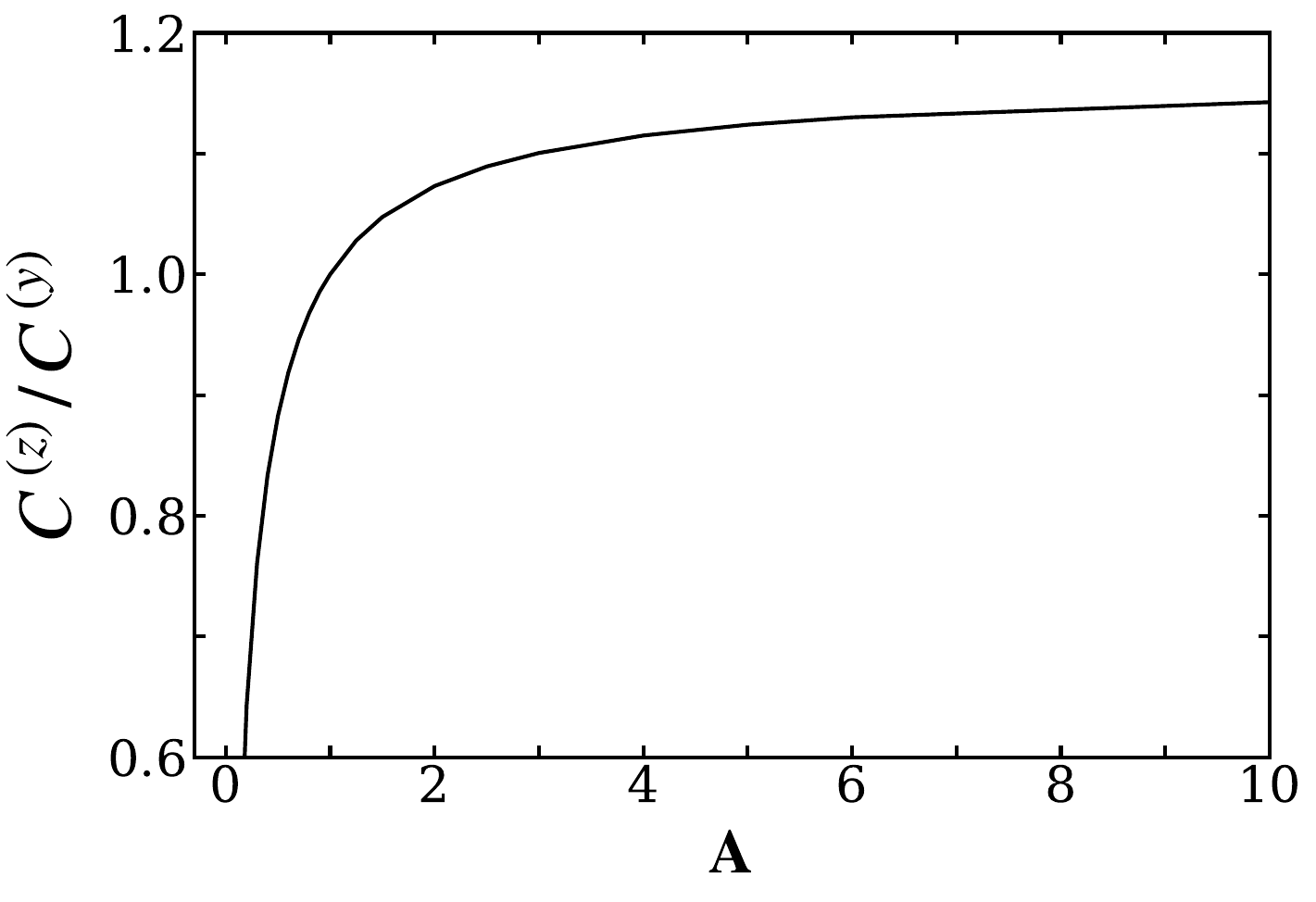} \caption{Left panel: Efimov scaling parameter as a function of the mass ratio $A = m_c/m_n$. Right panel: ratio between constant weights, $C^{(z)}/C^{(y)}$, of the Faddeev components, where $y$ and $z$ represent the components that are written in terms of the Jacobi coordinates that connects equal and different particles, respectively.} \label{Scalep} 
\end{figure}

Following the well-known behavior we remind through the left panel of Fig.~\ref{Scalep} that the Efimov scaling parameter $s_0$ increases slowly with the mass ratio for $A>1$, approaching the constant value, $s_0 \approx 1.14$, as $A\to\infty$, while for $A<1$ it grows rapidly and diverges as $A\to0$. This behavior is in agreement with that shown in Refs.~\cite{Braaten2006,Yamashita2013}. Since $s_0$ controls the discrete scaling of Efimov states, its increase, as known, implies a larger density of nodes of the wave function in a finite region. The right panel of Fig.~\ref{Scalep} shows the ratio between the weights of the Faddeev components, $C^{(z)}/C^{(y)}$. For $A>1$, this ratio follows a trend similar to that of $s_0$, approaching the constant value, $C^{(z)}/C^{(y)} \approx 1.16$, as $A\to\infty$. In this regime, the system consists of two light particles and a heavy core. The heavy particle becomes nearly static, and the exchange dynamics responsible for binding the Efimov state is dominated by the motion of the two light particles. As a consequence, the Faddeev component associated with the $Z$ Jacobi coordinates, which connect the two light particles through the heavy core, contributes slightly more than the $Y$ components. For $A<1$, the behavior is opposite: the ratio $C^{(z)}/C^{(y)}$ decreases and vanishes as $A\to0$. In this limit the system contains two heavy particles and a light one. The light particle moves rapidly between the heavy ones and mediates the effective interaction that binds the three-body system, making the $Y$ Jacobi components — which describe the distance between the light particle and each heavy particle — the dominant contributions.

\begin{figure}[ht!]
\centering
\includegraphics[width=6.38cm]{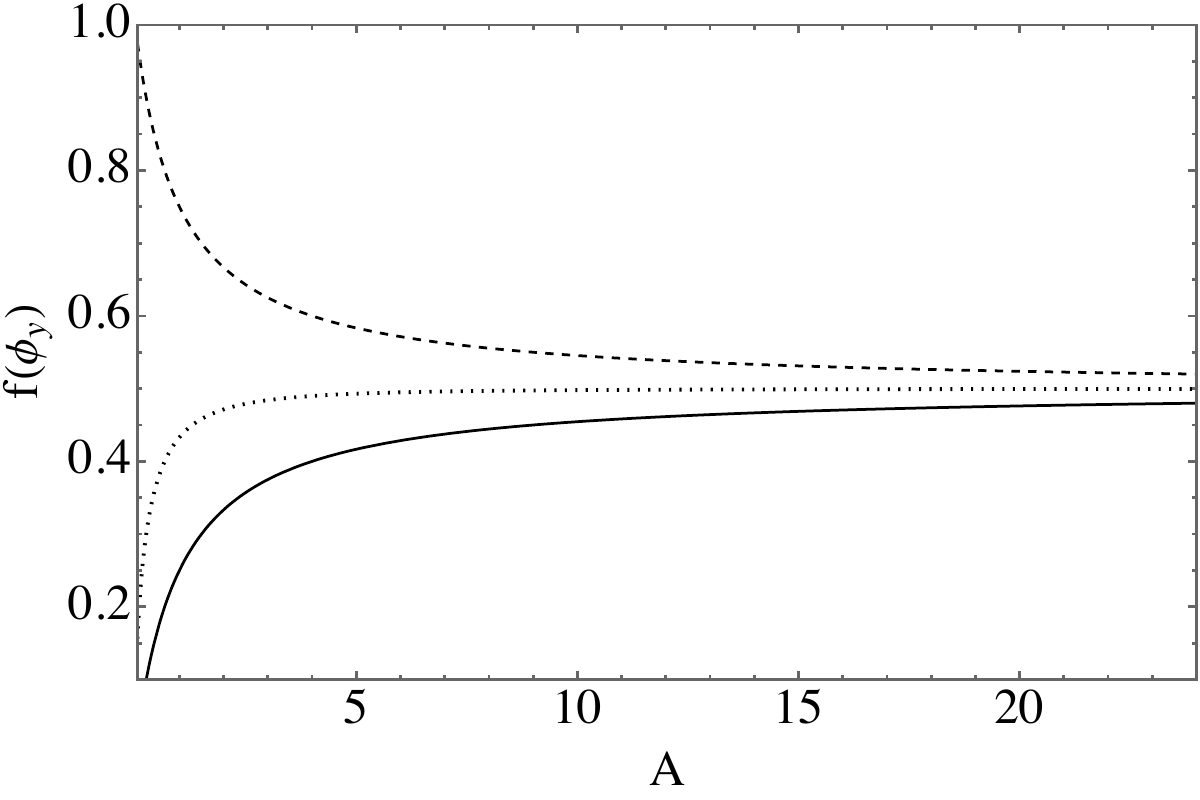} 
\centering
\includegraphics[width=6.28cm]{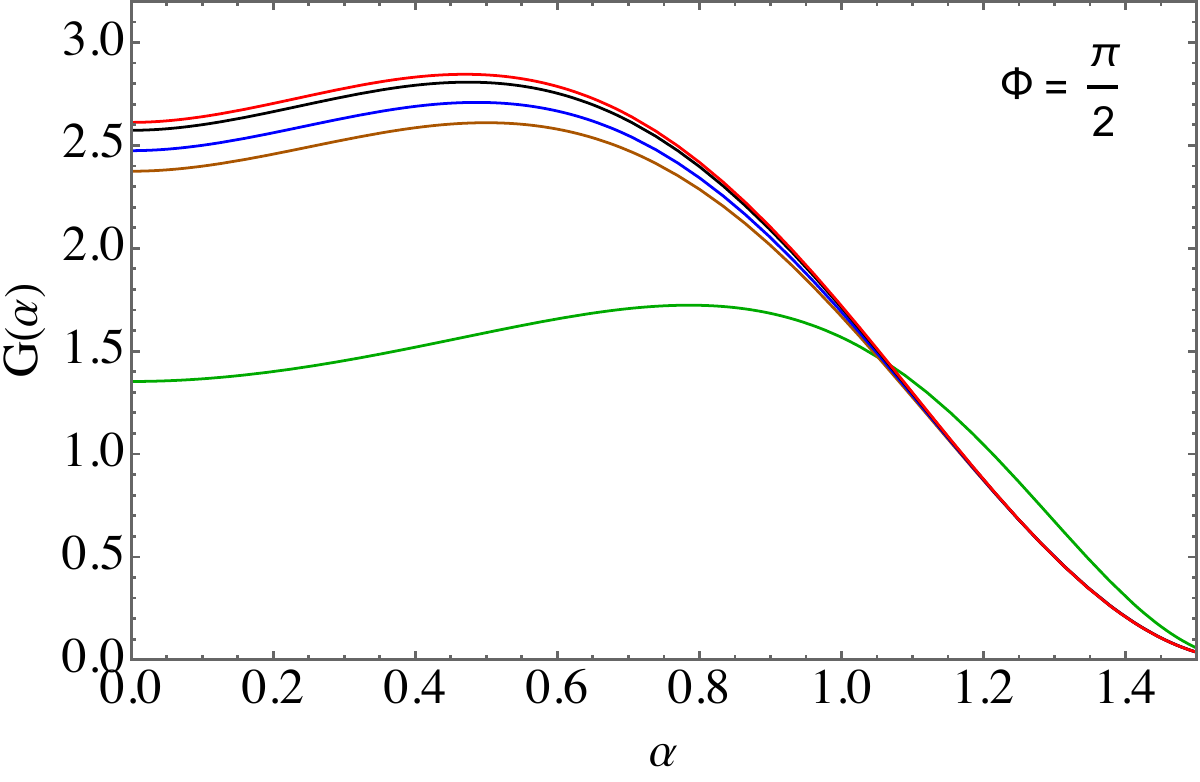} 
\centering
\includegraphics[width=6.28cm]{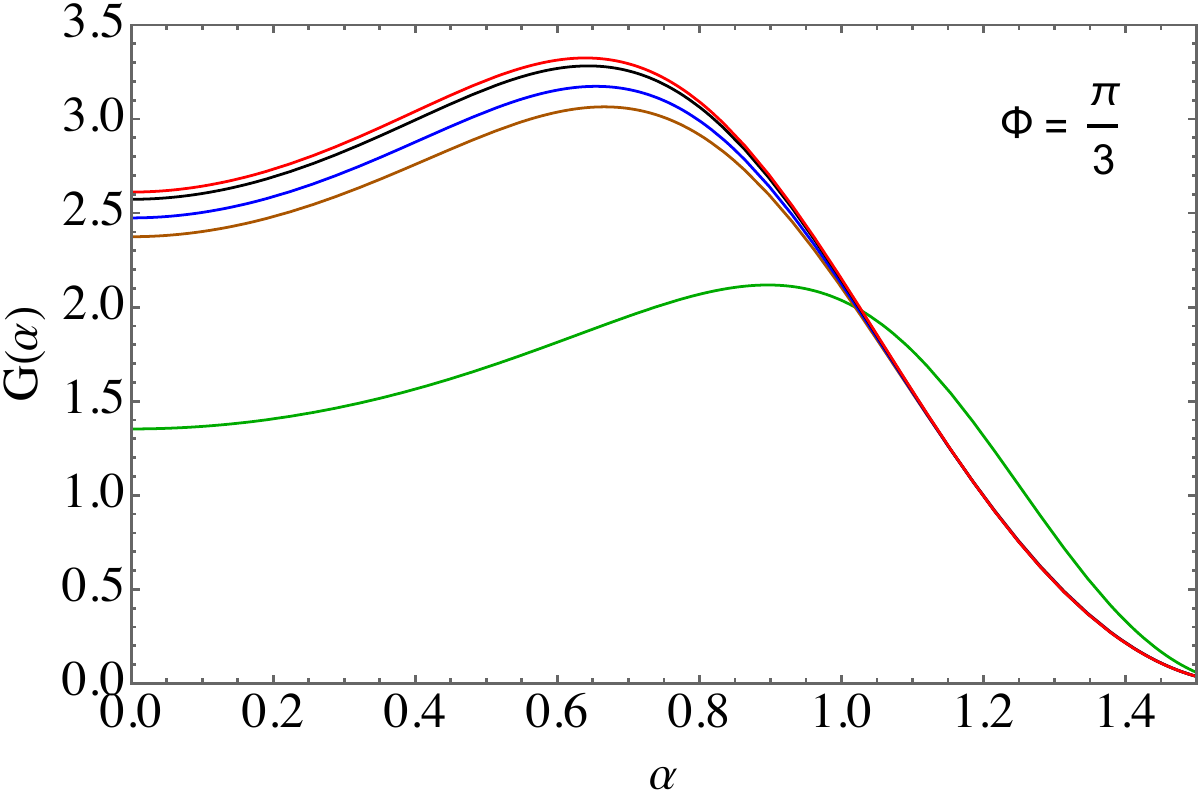}
\centering
\includegraphics[width=6.28cm]{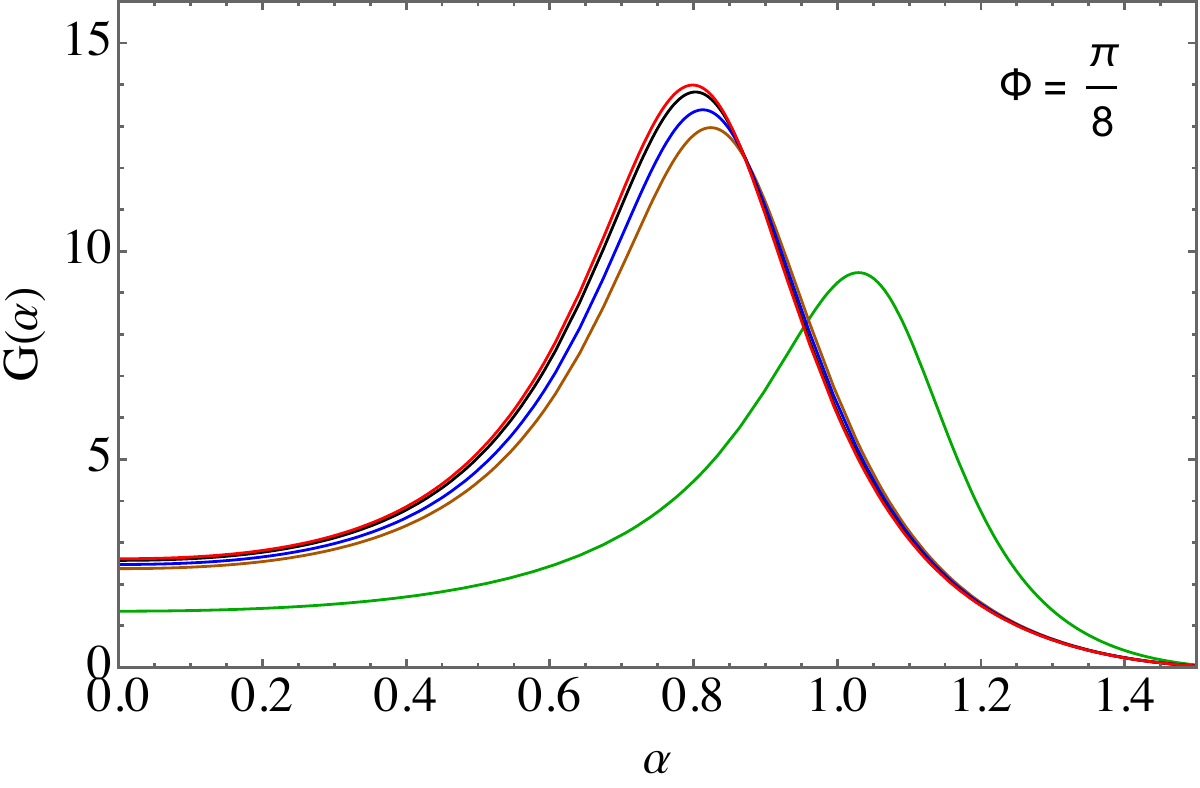}
\caption{Left upper panel: Transformation angle functions $\cos^2\phi_y$ (dashed line), $\sin^2\phi_y$ (continuous line) and $\cos\phi_y$$\sin\phi_y$ (dotted line) that appear inside the total hyperangular function (please, see Eq.~\eqref{totalwav1}) as functions of the mass ratio. Other panels: hyperangular wave function, $\sum_{i=1}^3 G(\alpha_i)$, for different angles ($\Phi$) between the Jacobi vectors, $\vec{r}'_z$ and  $\vec{\rho}'_z$, as functions of the hyperangle. Green, brown, blue, black and red curves for mass ratios $A = 1, 9, 12$, $17$ and $20$, respectively.} \label{angle_part} 
\end{figure}

The behavior of the scale parameter and of the Faddeev component weights for $A > 1$, where the mass configuration of halo nuclei systems lies, suggests a universal character in the internal structure of these systems. This can be seen by noting that the three-body wave function, Eq.~\eqref{eq:totalwavefunction}, exhibits a structure in which the hyperradial part encodes information about the spatial extent, while the angular part carries information about the internal configuration. As one can observe, the Faddeev component weights multiply the hyperangular part, which depends on $s_0$, on the ratio between the absolute value of the Jacobi coordinates, on the angles ($\Phi$) between the Jacobi vectors, $\vec{r}'_z$ and $\vec{\rho}'_z$, and on the transformation angles in Eq.~\eqref{transformationang}. In appendix~\ref{secA1}, we show explicitly the functions of the transformation angle that appear inside the hyperangular part of the three-body wave function, and, in the upper left panel of Fig.~\ref{angle_part}, we show these functions for $A = 1$, as well as for the mass configuration of the halo nuclei $^{11}\mathrm{Li}$, $^{14}\mathrm{Be}$, $^{19}\mathrm{B}$, and $^{22}\mathrm{C}$. As one can observe, all these functions also approach a constant value as $A$ increases. Associated to these results, one can think about the following situation: as $A$ increases, the system approaches a situation in which the core is a heavy particle that is approximately fixed in the space, while the neutrons move fast around it. This picture corroborates the universal geometry in the $A \gg 1$ regime. In the upper right and lower panels of Fig.~\ref{angle_part}, we show the hyperangular part of the wave function, $\sum_{i=1}^3 G(\alpha_i)$, for different angles ($\Phi$) between the Jacobi vectors, $\vec{r}'_z$ and  $\vec{\rho}'_z$, where a very similar shape for the halo nuclei mass ratios $A = 9, 12, 17$ and $20$, can be appreciated. 

\begin{figure}[ht!]
\centering
\includegraphics[width=6.53cm]{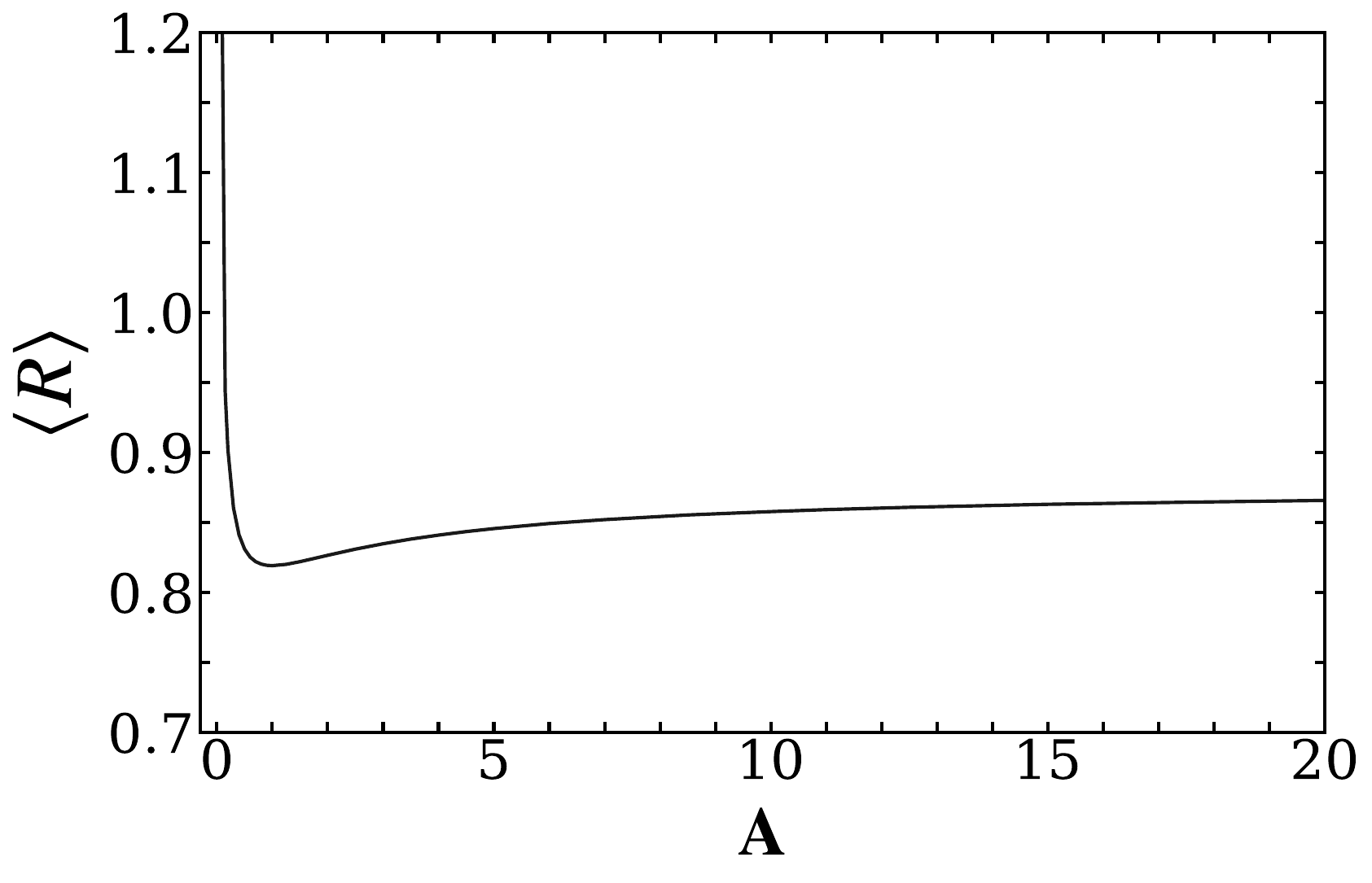}
\centering
\includegraphics[width=6.43cm]{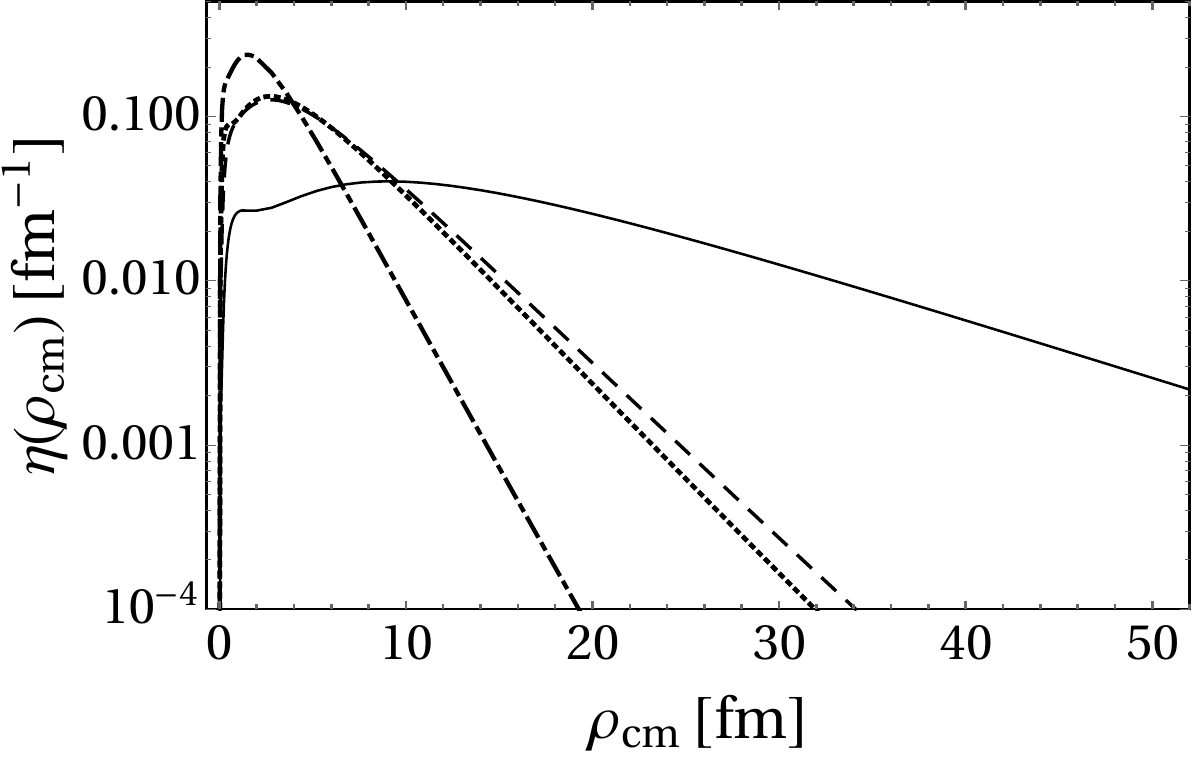} 
\caption{Left panel: dimensionless root-mean-square hyperradius, $\langle R \rangle \equiv \sqrt{|E_3|(\langle r'^2 \rangle + \langle \rho'^2 \rangle)}$, as a function of the mass ratio $A = m_c/m_n$. Right panel: Halo-neutron density, $\eta(\rho_{cm})$, with dashed-dotted, dotted, dashed and continuous lines for the halo-nuclei $^{14}$Be, $^{19}$B, $^{11}$Li and $^{22}$C, respectively.} \label{Ndensity} 
\end{figure}

In the left panel of Fig.~\ref{Ndensity}, we show the dimensionless root-mean-square hyperradius, $\langle R \rangle \equiv \sqrt{|E_3|(\langle r'^2 \rangle + \langle\rho'^2\rangle)}$, as a function of the mass ratio, $A$. This observable provides a notion of the overall size of the three-body system. Now, to analyze the halo character of the two-neutron halo systems \(^{11}\text{Li}\), \(^{14}\text{Be}\), \(^{19}\text{B}\),  and \(^{22}\text{C}\), in the right panel of Fig.~\ref{Ndensity}, we present the one-body density for the neutron, defined as
\begin{align}
    \eta(\rho_{cm}) = \rho_{cm}^2\int  drd\Phi  ~r^2 \sin\Phi       ~|\Psi(r,\rho_{cm})|^2\,,
\end{align}
where $ \rho_{cm} = \frac{A+1}{A+2}\rho\,$, is the distance between the neutron and the center of mass of the three-body system, while $\Phi$ is the angle between the $\vec{r}'_z$ and $\frac{A+1}{A+2}\vec{\rho}'_z$ vectors. To calculate these densities, we used the binding momentum corresponding to the two-neutron separation energies, $S_{2n}$, listed in Tab.~\ref{tab:neutron_properties}. For $^{22}$C, where the extracted $S_{2n}$ remains experimentally challenging and somewhat model dependent, while interaction-cross-section measurements provide direct
constraints on its spatial extension~\cite{Tanaka2010,Togano2016}. In this work, we adopt the value listed in NuDat/ENSDF as a representative input. As expected, the systems with smaller $S_{2n}$ values exhibit the most extended halos, being $^{22}$C the halo nucleus with the most prominent halo. 

\begin{table}[t!]
\centering
\caption{Two-neutron separation energies ($S_{2n}$) used for the two-neutron halo nuclei considered in this work.}
\label{tab:neutron_properties}
\begin{tabular}{l c c c}
\toprule
\textbf{System} & \textbf{Mass Ratio} $(A)$ & \textbf{$S_{2n}$ (MeV)} & \textbf{Source} \\
\midrule
$^{11}$Li & 9  & 0.3693 & \cite{NNDC2024} \\
$^{14}$Be & 12 & 1.270  & \cite{NNDC2024} \\
$^{19}$B  & 17 & 0.384  & \cite{Hiyama2022} \\
$^{22}$C  & 20 & 0.035  & \cite{NNDC2024} \\
\bottomrule
\end{tabular}
\end{table}

\subsection{Two-neutron halo nuclei geometry at the unitary limit}\label{subsec3.2}

In this subsection, we show results for average interparticle distances and separation angles as a function of the mass ratio, $A$. We also compute values for some of these quantities for the $^{11}$Li, $^{14}$Be, $^{19}$B and $^{22}$C halo nuclei. For $^{19}$B, the s-wave
dominance follows the three-body analysis of Hiyama et al.~\cite{Hiyama2022}.

\begin{figure}[t!]
\centering
\includegraphics[width=6.4cm]{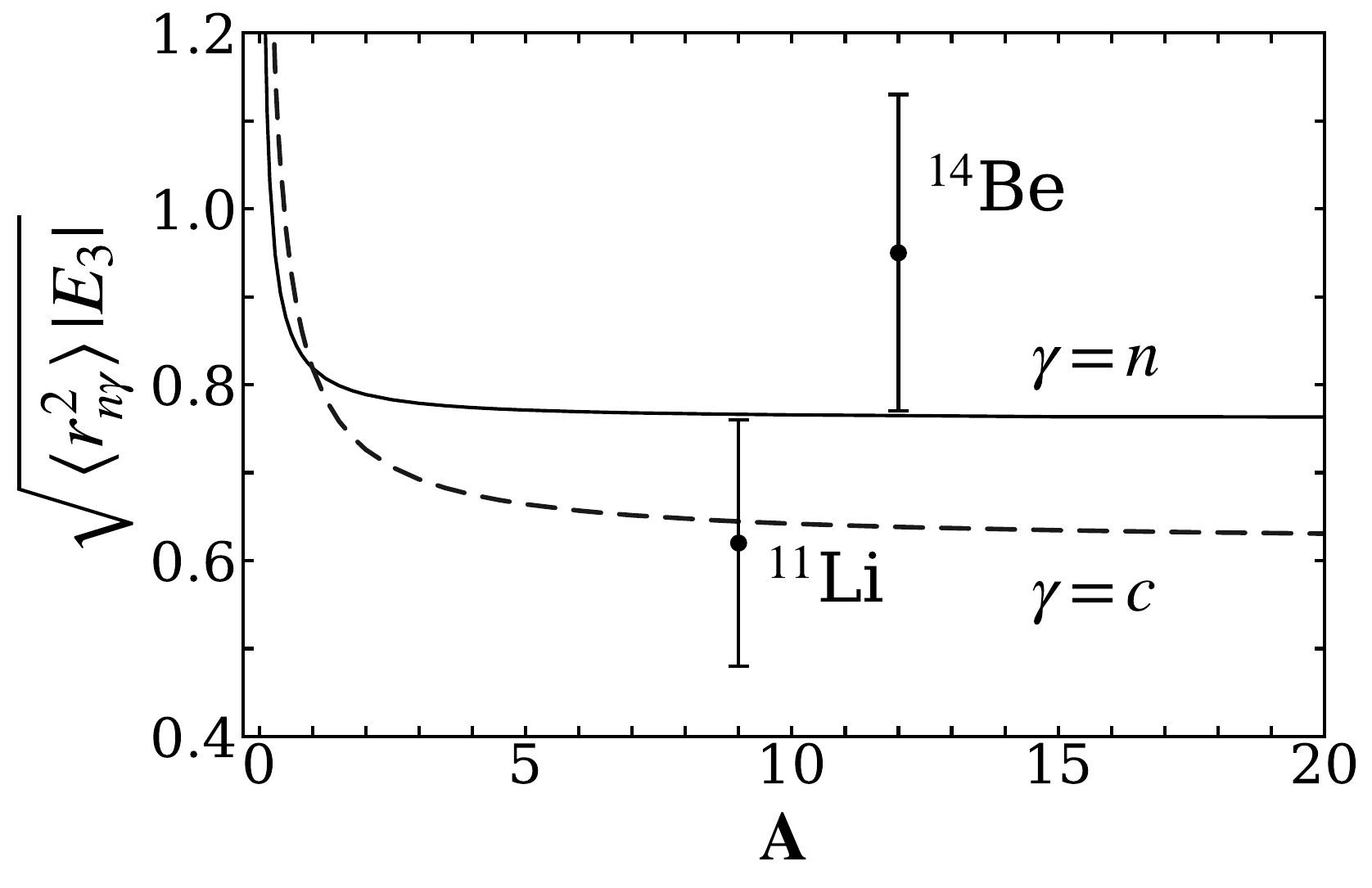}
\centering
\includegraphics[width=6.42cm]{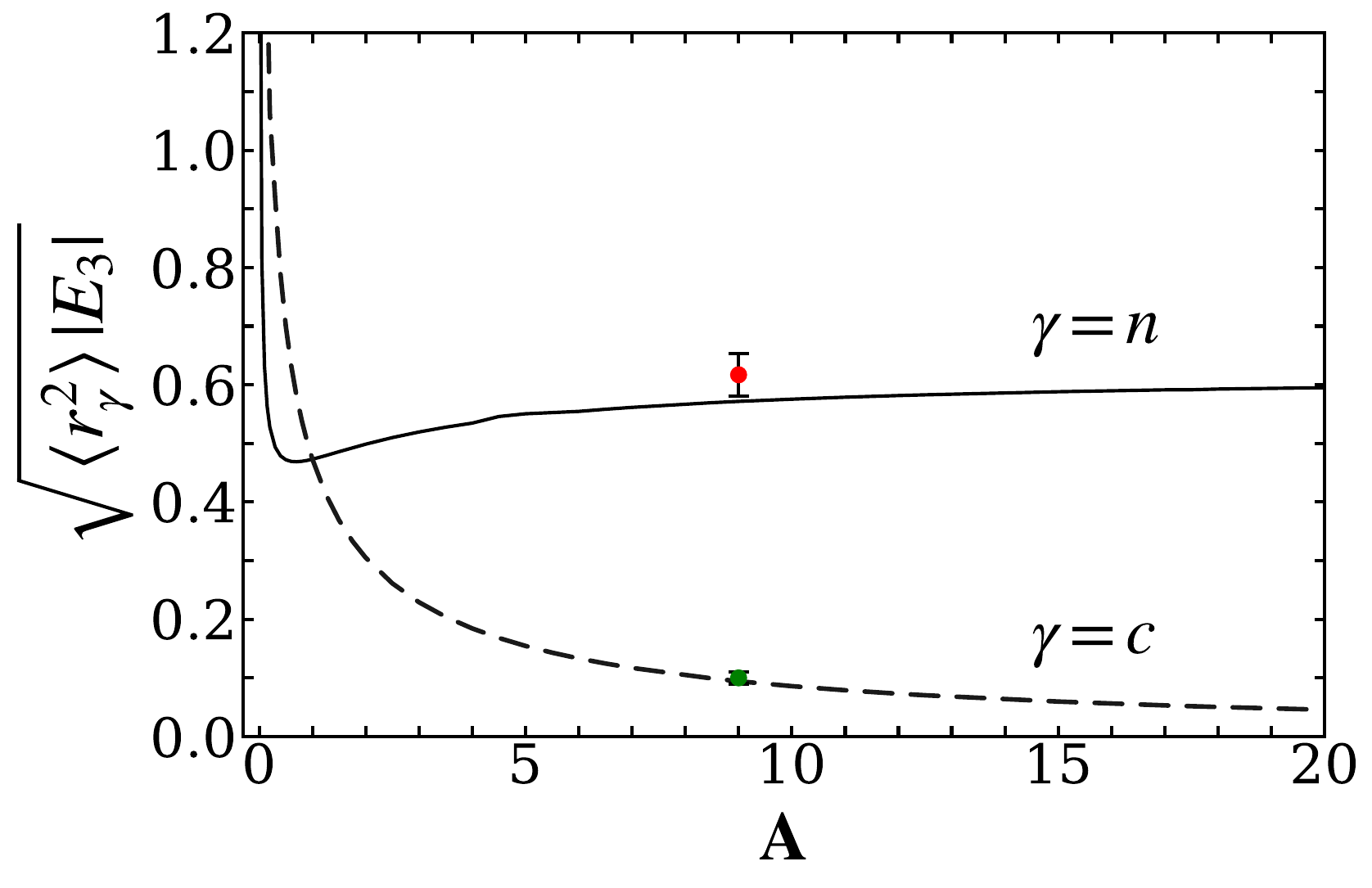} \caption{Dimensionless products $\sqrt{\langle r^2_{n\gamma} \rangle|E_3|}$ (left panel) and $\sqrt{\langle r^2_{\gamma} \rangle|E_3|}$ (right panel) with $\gamma \equiv \{n,c\}$ as functions of the mass ratio $A = m_c/m_n$ at the unitary limit. The
experimental data shown in the left panel are for $\sqrt{\langle r^2_{nn} \rangle|E_3|}$ and were reproduced from Refs.~\cite{Marques2000,Marques2001}. The experimental points for $^{11}$Li in the right panel were reproduced from Refs.~\cite{Egelhof2002} (red point for $\sqrt{\langle r^2_{n} \rangle |E_3|}$) and~\cite{Sanchez2006} (green point for $\sqrt{\langle r^2_{c} \rangle|E_3|}$). These curves were also obtained in Ref.~\cite{Frederico2012}, via numerical methods.} \label{fig3} 
\end{figure}

In Fig.~\ref{fig3}, we show the dimensionless root mean square distances, $\sqrt{\langle r^2_{n\gamma} \rangle|E_3|}$ (left panel) and $\sqrt{\langle r^2_{\gamma} \rangle|E_3|}$ (right panel), with $\gamma \equiv \{n,c\}$, as functions of the mass ratio $A = m_c/m_n$. All of the distances associated to these observables are depicted in Fig.~\ref{figangles}. In the left panel of Fig.~\ref{fig3}, for $A > 1$, we observe that $\sqrt{\langle r^2_{n\gamma}\rangle|E_3|}$ with $\gamma \equiv n$, which is associated to the neutron-neutron separation distance when $A > 1$, present larger values than the $\gamma \equiv c$ curve, which is associated to the separation distance between the core and the neutrons. Besides that, we see that both observables tend to constant values as $A \to \infty$, which is a consequence of the behavior of the hyperangular part of the three-body wave function that was discussed previously. For $A = 1$, as expected, the two curves present the same value and the root mean square distances of the three-body systems form an equilateral triangle. For $A < 1$, the $\gamma \equiv c$ curve starts to display larger values than the $\gamma \equiv n$ one, and both diverges as $A \to 0$. In the right panel of Fig.~\ref{fig3}, we observe behavior similar to that shown in the previous figure. For $A>1$, the dimensionless root-mean-square distance of the neutron from the three-body center of mass is larger than that of the core, with both approaching constant values as the mass ratio tends to infinity. For $A = 1$, these quantities become equivalent, while for $A < 1$ the position of the curves invert, with both diverging as the mass ratio goes to zero. In this figure we can also see a red and a green point, which correspond to experimental data obtained in Refs.~\cite{Egelhof2002} and \cite{Sanchez2006}, respectively. Also, it is important to mention that our curves, which were obtained by using an analytical three-body wave function, matches the numerically calculated data obtained in Ref.~\cite{Frederico2012}. Finally, in Tab.~\ref{tab:calculated_properties}, we show numerical values for the root-mean-square interparticle distances in femtometers for the $^{11}$Li, $^{14}$Be, $^{19}$B and $^{22}$C halo nuclei, where the values of $S_{2n}$ in Tab.~\ref{tab:neutron_properties} were used as the three-body binding energies for these halo systems. Our results exhibit significantly larger values than those obtained using finite two-body scattering lengths for the neutron-neutron and neutron-core systems~\cite{Frederico2012}. This trend is consistent with the universal behavior near the unitary limit. In this regime, the diverging two-body scattering length removes the finite scale that would otherwise constrain the system. The resulting scale invariance allows for the formation of extended Efimov states, whose spatial extent and resonant effects are maximized. 

\begin{table}[t!]
\centering
\caption{Calculated root-mean-square distances in femtometers. The distances are depicted in Fig.~\ref{figangles}.}
\label{tab:calculated_properties}
\small
\begin{tabular}{l c c c c}
\toprule
\textbf{System} &
\textbf{$\sqrt{\langle r^2_{nn} \rangle}$} &
\textbf{$\sqrt{\langle r^2_{nc} \rangle}$} &
\textbf{$\sqrt{\langle r^2_n \rangle}$} &
\textbf{$\sqrt{\langle r^2_c \rangle}$} \\
\midrule
$^{11}$Li & 8.12 & 6.83 & 6.06 & 1.00 \\
$^{14}$Be & 4.38 & 3.65 & 3.33 & 0.42 \\
$^{19}$B  & 7.93 & 6.57 & 6.14 & 0.55 \\
$^{22}$C\textsuperscript{(a)} & 26.27 & 21.70 & 20.47 & 1.57 \\
\bottomrule
\end{tabular}

\vspace{0.15cm}

\parbox{0.92\linewidth}{\footnotesize
(a) The \(^{22}\mathrm C\) distances are especially sensitive to the adopted
\(S_{2n}\), since rms distances scale approximately as
\(\sqrt{\langle r^2\rangle}\sim 1/\sqrt{S_{2n}}\). Therefore, the values quoted here
should be read in connection with the chosen \(S_{2n}=0.035\) MeV.}

\end{table}

\begin{figure}[ht!]
\centering
\includegraphics[width=6.4cm]{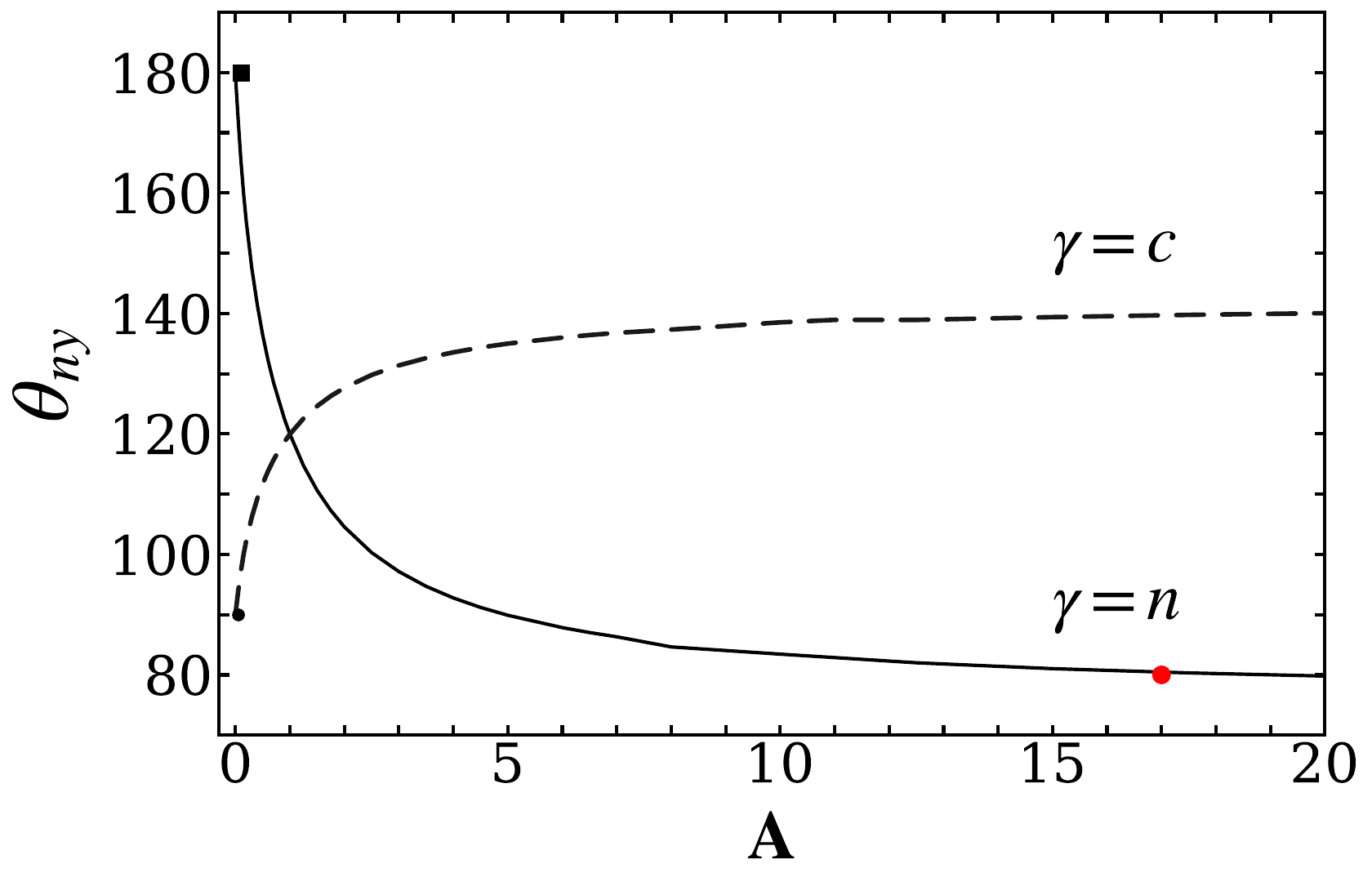}
\centering
\includegraphics[width=6.4cm]{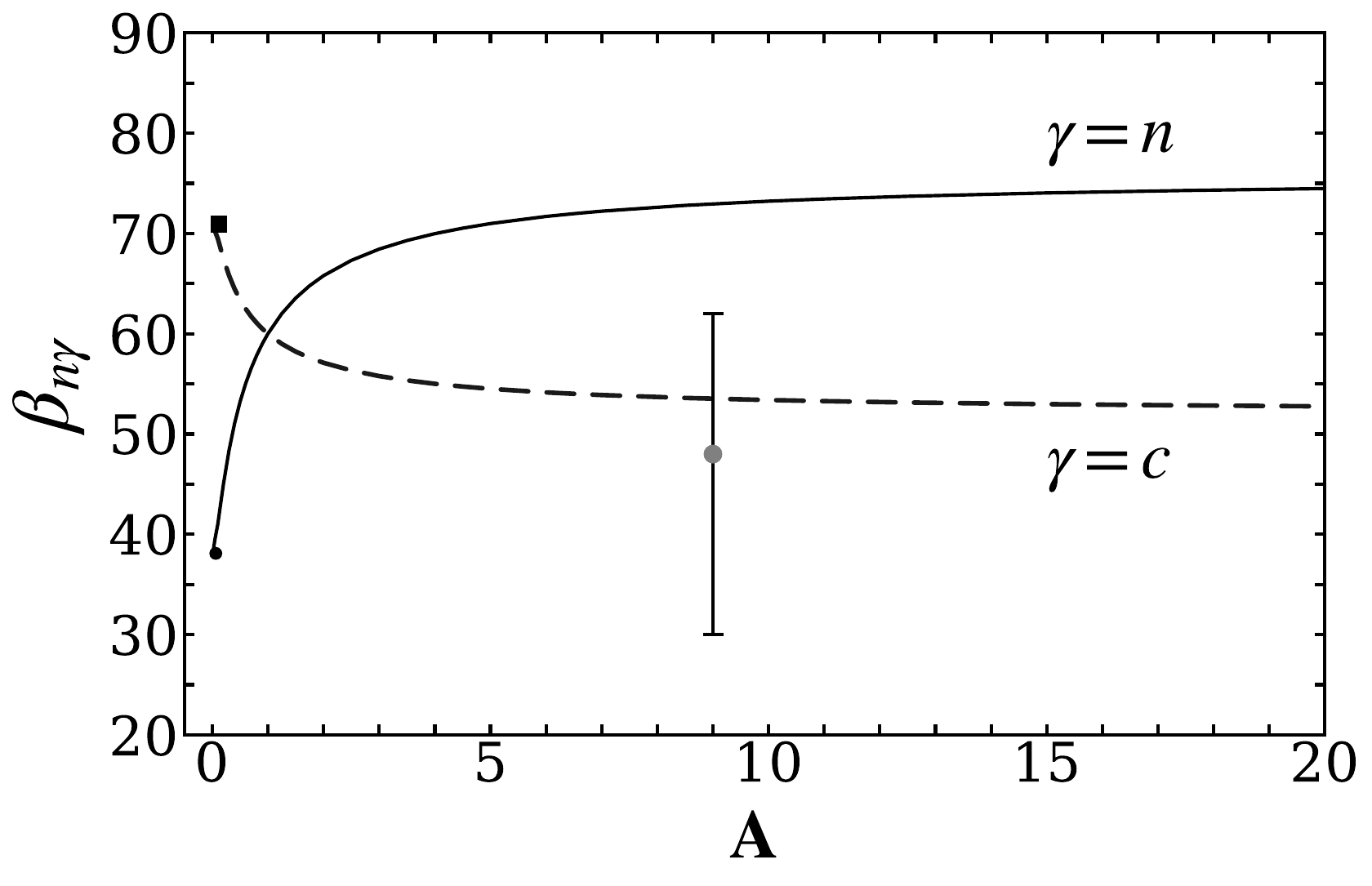}
\caption{Average angles $\theta_{n\gamma} (\gamma\equiv {n,c})$ (left panel) and $\beta_{n\gamma} (\gamma\equiv {n,c})$ (right panel) as functions of the mass ratio $A = m_c/m_n$. The red point in the left panel indicates the value for $\theta_{nn}$ obtained in Ref.~\cite{Hiyama2022} with different models for neutron-neutron and neutron-core interactions. The gray point in right panel indicates the experimental value for $\beta_{nn}$ obtained in Ref.~\cite{Nakamura2006}.} \label{fig4} 
\end{figure}

The adoption of the unitary limit in the calculation of the observables
reported in Tab.~\ref{tab:calculated_properties} introduces a systematic
offset with respect to more realistic finite-range calculations. This offset is
associated mainly with the two effects neglected at unitarity: finite two-body
scattering lengths and finite effective ranges. The size of these corrections is
controlled by the dimensionless products between those quantities and the three-body
binding momentum ($\kappa_0 = \sqrt{-E_3}$ in units of $\hbar=m_n=1$). For example, the neutron-neutron scattering length has a finite value of $a_{nn}=-18.7$~fm. The corresponding effective-range correction is
of order $r_0/a_{nn}\simeq 8\%$, while the finite-scattering-length correction,
$1/(\kappa_0 a_{nn})$, is typically of the order of a few tenths for the more
strongly bound halo systems. 

Explicit halo-EFT calculations by Canham and Hammer~\cite{Canham2008,Canham2010}, performed at the physical
scattering lengths, show that corrections of this size translate into shifts of
the rms radii of a few percent up to roughly fifteen percent for the Borromean
systems $^{11}$Li and $^{14}$Be, while the inclusion of finite-range effects
substantially reduces the theoretical uncertainty bands. The present calculation  is performed at the strict unitarity limit, and should
therefore be interpreted as a universal baseline rather than as a precision
description of each nucleus.

A useful comparison can also be made with the calculation of
Naidon~\cite{Naidon2023}, which includes finite two-body input for
$^{11}$Li. In that work, the quoted rms distance
$\sqrt{\langle R_{12,3}^{2}\rangle}=4.70$~fm corresponds to the distance between
the core and the center of mass of the two-neutron subsystem. To obtain the rms
distance between the core and the center of mass of the full three-body system,
we use the relation
\begin{equation}
    \mathbf r_{c}
    =
    \frac{2}{A+2}\mathbf R_{12,3}.
\end{equation}
For $^{11}$Li, treated as $^{9}{\rm Li}+n+n$, one has $A=9$, and therefore
\begin{equation}
    \sqrt{\langle r_c^2\rangle}
    =
    \frac{2}{11}
    \sqrt{\langle R_{12,3}^{2}\rangle}
    \simeq
    0.85~{\rm fm}.
\end{equation}
This should be compared with the unitary-limit value
$\sqrt{\langle r_c^2\rangle}=1.00$~fm reported in
Tab.~\ref{tab:calculated_properties}. The difference, of order $15$--$20\%$, is
consistent with the expected size of finite-scattering-length and finite-range
corrections in $^{11}$Li.

Finally, since rms distances scale approximately with the inverse binding
momentum,
\begin{equation}
    \sqrt{\langle r^2\rangle}\sim \frac{1}{\sqrt{S_{2n}}},
\end{equation}
the values reported in Tab.~\ref{tab:calculated_properties} are 
sensitive to the adopted value of $S_{2n}$, which is relevant for the $^{22}$C halo properties. This sensitivity is crucial to determine $S_{2n}[^{22}\text{C}]$, which is not well constrained experimentally, from the matter-radius extractions reported
by Tanaka et al.~\cite{Tanaka2010} and Togano et al.~\cite{Togano2016}.

In Fig.~\ref{fig4}, we show the mean opening angles depicted in Fig.~\ref{figangles} as functions of the mass ratio, $A$. In the left panel, we can observe the behavior of the mean opening angles $\theta_{n\gamma} (\gamma\equiv {n,c})$, defined at the center of mass of the three-body system. As the mass ratio goes to zero, $\theta_{nn}$ goes to $180^\circ$, while $\theta_{nc}$ goes to $90^\circ$. This can be understood by thinking in how the three-body center of mass changes as two particles becomes much heavier than the third one. For $A \to 0$, the center of mass approaches the line that connects the two heavy particles, so that $\theta_{nn}$ approaches its maximum value and the $r_c$ becomes perpendicular to $r_n$. For $A = 1$, these mean angles are equal, corroborating the equilateral triangle form. For $A >1$, we have that $\theta_{nc} > \theta_{nn}$, with these angles approaching constant values as $A \to \infty$. In the right panel of Fig.~\ref{fig4}, we present the average opening angles $\beta_{n\gamma} (\gamma\equiv {n,c})$. These angles are defined at the center of the particles that compose the system. In this case, for $A > 1$, we have that $\beta_{nn} > \beta_{nc}$, with $\beta_{nn}$ and $\beta_{nc}$ approaching constant values as the mass ratio goes to infinity. This stems from the fact that in this mass configuration regime, $\sqrt{\frac{\langle r^2_{nn}\rangle}{\langle r^2_{nc} \rangle}} > 1$, approaching a constant value as $A \to \infty$. It is important to note that in the limit of large mass ratio, the center of mass goes up and reaches the center of the core for $A \to \infty$, so that in this limit we have $\theta_{nn} = \beta_{nn}$. For $A = 1$, as expected, we have $\beta_{nn} = \beta_{nc} = 60^\circ$. For $A < 1$, we have $\sqrt{\frac{\langle r^2_{nn} \rangle}{\langle r^2_{nc} \rangle}} < 1$, so that in this regime $\beta_{nc}$ becomes larger than $\beta_{nn}$. 

\section{Conclusion}\label{sec4}

Using the analytical three-body wave function at the unitary limit, derived through the Faddeev decomposition together with the Bethe–Peierls boundary condition, we investigated the geometric and structural properties of the halo in light exotic two-neutron halo nuclei  dominated by two-body 
s-wave interactions. In particular, we analyzed 
$^{11}$Li, $^{14}$Be, $^{19}$B and $^{22}$C, treating them as neutron–neutron–core three-body systems within the framework of Efimov physics. In this way, the only physical scale is the two-neutron separation energies that corresponds to the three-body binding energies of these Efimov-like states. The analytic formulation valid for arbitrary mass ratios enabled the calculation of probability densities, root-mean-square interparticle distances and opening angles, providing a unified description of their geometric configurations at unitarity.

Our results indicate that these observables rapidly converge to constant asymptotic values as the mass of the core increases relative to the neutron mass. This behavior reflects the emergence of universal geometry in strongly mass-imbalanced halo systems and the dominance of the light-particle correlations near unitarity. The comparison with the limited available
experimental data is consistent with the unitary-limit prediction for some observables, but reveals significant tensions for others.

In future works, we aim to extend the present analysis by incorporating corrections beyond the unitarity, including the effects from the finite scattering lengths, effective ranges, and possible
p-wave contributions. We expect to improve the quantitative agreement with experimental data and to shed further light on the limits of universality and the role of nonuniversal effects in neutron-rich halo nuclei.

\section{Acknowledgements}

This study was financed, in part, by the Fundação de Amparo à Pesquisa do Estado de São Paulo (FAPESP), Brazil [grant numbers 2023/13749-1 (T.F.), 2024/17816-8 (T.F. and M.T.Y.), 2025/05312-8 (T.F. and M.T.Y.), 2023/08600-9 and 2025/15267-0 (R.M.F. and T.F.)] and by the Conselho Nacional de Desenvolvimento 
Cient\'{i}fico e Tecnol\'{o}gico (CNPq) [grant numbers 306834/2022-7 (T.F.), 151403/2025-2 (D.S.R.) and  
302105/2022-0 (M.T.Y.)]. This work is a part of the
project Instituto Nacional de  Ci\^{e}ncia e Tecnologia - F\'{\i}sica
Nuclear e Aplica\c{c}\~{o}es  Proc. No. CNPq 408419/2024-5.

\begin{appendices}

\section{Three-body wave function for a YYZ system}\label{secA1}

In this appendix, the three-body wave function in Eq.~\eqref{eq:totalwavefunction} is expressed in terms of one pair of Jacobi coordinates for a system that consists of two particles with identical masses and a third particle with a different mass value, namely YYZ system. We choose to express the wave function in terms of the so called T-Jacobi coordinates: the $|\vec{r_z'}| \equiv r$ coordinate, namely, the distance between the identical particles (YY) and $|\vec{\rho_z'}|$ the distance between the different particle (Z) and the center of mass of the YY subsystem. In order for the three-body wave function to be written in terms of just one pair of coordinates, we have to express the coordinates $\vec{|r_y'}|$ and $|\vec{\rho_y'}|$ in terms of the Z-coordinates. This can be done by using Eq.~\eqref{transformationang}, which can be expressed in the YYZ case as
\begin{align}
    \phi_y = \arctan\left(\sqrt{\frac{2 + A}{A}}\right).
\end{align}
 In this way, the three-body wave function at the unitary limit in terms of the T-Jacobi coordinates is written as
\begin{eqnarray}
\Psi(r',\rho',\Phi) &= & N_{\Psi}\frac{K_{\mathrm{i} s_0}\!\left(\sqrt{2}\,\kappa_0 \sqrt{r'^2 + \rho'^2}\right)}{(r'^2+\rho'^2)} \, \left[\frac{C^{(z)}}{C^{(y)}}\frac{\sin\!\left[\mathrm{i} s_0 \!\left(\frac{\pi}{2} - \arctan\!\frac{r'}{\rho'}\right)\right]}{\,\sin\!\left[2\,\arctan\!\frac{r'}{\rho'}\right]}\right.\nonumber \\
&+&
\frac{
\sin\!\left[\mathrm{i} s_0 \!\left(\frac{\pi}{2} -
\arctan\sqrt{
\frac{
r'^2\cos^2\phi_y + \rho'^2\sin^2\phi_y - 2 r'\rho' \cos\phi_y \sin\phi_y \cos\Phi
}{
r'^2\sin^2\phi_y + \rho'^2\cos^2\phi_y + 2 r'\rho' \cos\phi_y \sin\phi_y \cos\Phi
}
}\right)\right]
}{\sin\!\left[
2\,\arctan \sqrt{
\frac{
r'^2\cos^2\phi_y + \rho'^2\sin^2\phi_y - 2 r'\rho' \cos\phi_y \sin\phi_y \cos\Phi
}{
r'^2\sin^2\phi_y + \rho'^2\cos^2\phi_y + 2 r'\rho' \cos\phi_y \sin\phi_y \cos\Phi
}
}\right]
}
\nonumber \\
&+&\left.
\frac{
\sin\!\left[\mathrm{i} s_0 \!\left(\frac{\pi}{2} -
\arctan\sqrt{
\frac{
r'^2\cos^2\phi_y + \rho'^2\sin^2\phi_y + 2 r'\rho' \cos\phi_y \sin\phi_y \cos\Phi
}{
r'^2\sin^2\phi_y + \rho'^2\cos^2\phi_y - 2 r'\rho' \cos\phi_y \sin\phi_y \cos\Phi
}
}\right)\right]
}{
\sin\!\left[
2\,\arctan\sqrt{
\frac{
r'^2\cos^2\phi_y + \rho'^2\sin^2\phi_y + 2 r'\rho' \cos\phi_y \sin\phi_y \cos\Phi
}{
r'^2\sin^2\phi_y + \rho'^2\cos^2\phi_y - 2 r'\rho' \cos\phi_y \sin\phi_y \cos\Phi
}
}\right]
}\right]\,,\hspace{0.7cm}
\label{totalwav1}
\end{eqnarray}
 where $\Phi$ is the angle between the vectors of the Z-type Jacobi coordinates.
The coefficient $C^{(z)}$ corresponds to the weight of the Faddeev component
originally written in the Z Jacobi set, while $C^{(y)}$ is the weight of the
two components associated with the Y Jacobi sets. In Eq.~\eqref{totalwav1},
all three Faddeev components are expressed in the Z (or T) coordinate set.
The two Y components are obtained through the transformation of the Jacobi
coordinates. For convenience, we factor out the Y weight and absorb the
overall factor $C^{(y)}$ into the normalization constant.

\section{Faddeev components weights and scale parameter}\label{secA2}

In this appendix, we show the equations that we use to obtain the weights of the Faddeev components and the scale parameter. 
In three dimensions, for the YYZ system, the BP leads to a set of coupled equations that can be written in matrix form as
\begin{equation}
    \begin{bmatrix}
    \frac{\sin\left(\mathrm{i} s_0\left(\frac{\pi}{2} - \phi_z\right)\right)}{ \sin\phi_z\cos \phi_z}  -\mathrm{i} s_0\cos\left(\mathrm{i} s_0\frac{\pi}{2}\right)
    & \frac{C^{(z)}}{C^{(y)}}\frac{\sin\left(\mathrm{i} s_0\left(\frac{\pi}{2} - \phi_y\right)\right)}{\sin \phi_y\cos \phi_y} \\
    2\frac{\sin\left(\mathrm{i} s_0\left(\frac{\pi}{2} - \phi_y \right)\right)}{\sin \phi_y \cos \phi_y} & -\frac{C^{(z)}}{C^{(y)}}\mathrm{i} s_0\cos\left(\mathrm{i} s_0\frac{\pi}{2}\right)
\end{bmatrix}
\begin{bmatrix}
    F^{(y)}(R) \\
    F^{(z)}(R) 
\end{bmatrix}
= 0\,.
\label{matrix}
\end{equation}

The Efimov scaling parameter can be obtained by imposing that the determinant of the matrix has to vanish, which is the condition imposed to the homogeneous Eq.~\eqref{matrix} to have a non-trivial solution. The ratio between the weights of the Faddeev components can be obtained by solving one of the lines of Eq.~\eqref{matrix} as
\begin{align}
    \frac{C^{(z)}}{C^{(y)}} =\left[ \frac{\sin\left(\mathrm{i} s_0\left(\frac{\pi}{2} - \phi_z\right)\right)}{ \sin\phi_z\cos \phi_z} -\mathrm{i} s_0\cos\left(\mathrm{i} s_0\frac{\pi}{2}\right)\right] \left[-\frac{\sin\left(\mathrm{i} s_0\left(\frac{\pi}{2} - \phi_y\right)\right)}{\sin \phi_y\cos \phi_y}\right]^{-1}.
\end{align}

\end{appendices}

\end{document}